\documentclass[useAMS,usenatbib,usegraphicx]{mn2e}

\usepackage{mathptmx,subfigure}
\usepackage[fleqn]{amsmath}

\newcommand{\Alfven}{Alfv\'en}
\newcommand{\Alfvenic}{Alfv\'enic}
\newcommand{\para}{\parallel}
\newcommand{\degrees}{\ensuremath{^\circ}}

\newcommand{\prl}{Phys. Rev. Lett.}
\newcommand{\pre}{Phys. Rev. E}
\newcommand{\apjl}{ApJ}
\newcommand{\apj}{ApJ}
\newcommand{\grl}{Geophys. Res. Lett.}
\newcommand{\jgr}{J. Geophys. Res.}
\newcommand{\apjs}{ApJS}
\newcommand{\ppcf}{Plasma Phys. Control. Fusion}

\newcommand{\pof}{Phys. Fluids}
\newcommand{\pop}{Phys. Plasmas}
\newcommand{\lrsp}{Living Rev. Solar Phys.}
\newcommand{\ang}{Ann. Geophys.}

\newcommand{\jopp}{J. Plasma Phys.}
\newcommand{\rslpsa}{Proc. R. Soc. Lond. Ser. A}
\newcommand{\araa}{ARA\&A}

\newcommand{\jfm}{J. Fluid Mech.}

\newcommand{\aap}{A\&A}

\newcommand{\jpp}{J. Plasma Phys.}
\newcommand{\mnras}{MNRAS}
\newcommand{\pra}{Phys. Rev. A}

\title[Anisotropy of \Alfvenic\ turbulence]{Anisotropy of \Alfvenic\ turbulence in the solar wind and numerical simulations}
\author[C. H. K. Chen et al.]{C. H. K. Chen,$^{1,2}$\thanks{E-mail:
chen@ssl.berkeley.edu} A. Mallet,$^{3}$ T. A. Yousef,$^{4}$ A. A. Schekochihin$^{3}$ and T. S. Horbury$^{2}$\\
$^{1}$Space Sciences Laboratory, University of California, Berkeley, CA 94720, USA\\
$^{2}$The Blackett Laboratory, Imperial College London, London SW7 2AZ\\
$^{3}$Rudolf Peierls Centre for Theoretical Physics, University of Oxford, Oxford OX1 3NP\\
$^{4}$Bj\o rn Stallares gt 1, 7042 Trondheim, Norway}

\begin{document}
\date{}
\pagerange{}
\pubyear{2011}
\maketitle

\begin{abstract}
We investigate the anisotropy of \Alfvenic\ turbulence in the inertial range of slow solar wind and in both driven and decaying reduced magnetohydrodynamic simulations. A direct comparison is made by measuring the anisotropic second-order structure functions in both data sets. In the solar wind, the perpendicular spectral index of the magnetic field is close to $-5/3$. In the forced simulation, it is close to $-5/3$ for the velocity and $-3/2$ for the magnetic field. In the decaying simulation, it is $-5/3$ for both fields. The spectral index becomes steeper at small angles to the local magnetic field direction in all cases. We also show that when using the global rather than local mean field, the anisotropic scaling of the simulations cannot always be properly measured.
\end{abstract}

\section{Introduction}

The solar wind is a turbulent plasma \citep[see reviews by][]{goldstein95a,bruno05a,horbury05} with a power spectrum extending over many orders of magnitude \citep[e.g.][]{coleman68}. Scales larger than the ion gyroradius are known as the inertial range, and the spectral indices at 1 AU are observed to be close to $-5/3$ for the magnetic and electric fields and $-3/2$ for velocity \citep[e.g.][]{matthaeus82a,bale05,podesta07a,tessein09,podesta10d}. There is also evidence that the fluctuations are predominantly \Alfvenic\ \citep[e.g.][]{belcher71,horbury95,bale05}.

Solar wind turbulence is anisotropic with respect to the direction of the magnetic field. For example, the magnetic field correlation length has been shown to vary depending on the angle of observation with respect to the field direction \citep{crooker82,matthaeus90,dasso05,osman07,weygand09}. The magnetic field power and spectral indices are also observed to be anisotropic: power at a fixed scale increases with angle to the magnetic field \citep{bieber96,horbury98b,osman09a} and the spectral index varies from $-2$ at small angles to between $-3/2$ and $-5/3$ in the field perpendicular direction \citep{horbury08,podesta09a,luo10,wicks10a,wicks11}. These observations are consistent with theories of critically balanced magnetohydrodynamic (MHD) turbulence, for example, that of \citet{goldreich95}, which predicts anisotropic fluctuations ($k_\perp>k_\para$) and the $-2$ and $-5/3$ spectral indices.

Many simulations of plasma turbulence have been performed, most of which have used the equations of incompressible MHD. When a strong mean magnetic field is present, the spectral index of the total energy is closer to $-3/2$ than $-5/3$ \citep{maron01,muller03,muller05,mason08,perez08,grappin10}, although a limited inertial range and the bottleneck effect \citep{falkovich94} make this number hard to determine precisely \citep{beresnyak09b,beresnyak11}.

Anisotropy has also been measured in MHD simulations. Early 2D simulations showed that the turbulence develops wavevector anisotropy so that the fluctuations have $k_\perp>k_\para$ \citep{shebalin83}, and this was later confirmed in 3D simulations \citep{oughton94,matthaeus96,milano01}. The anisotropy was found to be scale-dependent, such that $k_\para\sim k_\perp^{2/3}$ \citep{cho00,maron01}, in agreement with the critical balance predictions \citep{goldreich95}. An important point noted in these studies, and also in solar wind measurements \citep{horbury08}, was that the anisotropic scaling is with respect to the scale-dependent local mean field and not the global mean field.

The theory of \citet{goldreich95} was modified by \cite{boldyrev06} by including a phenomenon called scale-dependent dynamic alignment. In this theory, the velocity and magnetic field fluctuations align to within a smaller angle at smaller scales and the perpendicular spectral index becomes $-3/2$. There is evidence for this scale-dependent dynamic alignment in the solar wind \citep{podesta09e} and some driven MHD simulations \citep{mason06,mason08}, although higher resolution simulations suggest that the alignment saturates at small scales \citep{beresnyak11}.

To date, there has not been a measurement of the spectral index parallel to the local magnetic field in simulations. Measurements of the perpendicular spectral index in the solar wind and in simulations are also not always in agreement. It is important to be sure that the same quantities are being measured in both the solar wind and simulations and the subject of this paper is such a comparative study. We apply a similar analysis technique to both solar wind data and reduced MHD (RMHD) simulations, to make a direct comparison of the anisotropic scaling. In Section \ref{sec:sw} we present the solar wind analysis, in Section \ref{sec:sim} we present the simulation analysis, in Section \ref{sec:localvsglobal} we compare the local and global mean field methods and in Section \ref{sec:conc} we present our conclusions.

\section{Inertial Range Solar Wind Measurements}
\label{sec:sw}

\subsection{Data Intervals}
\label{sec:swintervals}

In this section, we apply the multispacecraft method of \citet{chen10b} to obtain the power and spectral index anisotropy of inertial range turbulence in the slow solar wind at 1 AU. The technique is applied to 65 1-hour intervals of data from the Cluster spacecraft \citep{escoubet01} from December 2005 to April 2006, when the typical separation between the four spacecraft was $\sim$ 10,000 km. The selected intervals are from the parts of the Cluster orbit where the spacecraft were in the free solar wind upstream of the bow shock at geocentric distances of between 15 $R_E$ and 20 $R_E$. They contain no evidence of ion foreshock activity: signatures typical of the ion foreshock, such as enhanced magnetic field fluctuations and high-energy ions, are not present. The time series were also inspected visually to ensure that they are approximately stationary and do not contain shocks or magnetic clouds. 

In the analysis, we use 4 s measurements of the magnetic field from the fluxgate magnetometer (FGM) \citep{balogh01} and velocity and density moments from the Cluster ion spectrometer (CIS) \citep{reme01}. The mean values of various parameters for the 65 intervals are given in Table \ref{tab:parameters}. The geometric mean is used for the ion beta, temperature anisotropy, gyroradius and \Alfven\ ratio. The intervals are in slow solar wind with a speed $<$ 550 km s$^{-1}$.

\begin{table}
\caption{Mean parameter values for the 65 solar wind intervals}
\label{tab:parameters}
\begin{tabular}{cc}
\hline
Solar wind speed ($v_{\text{sw}}$) & 360 $\pm$ 10 km s$^{-1}$ \\
Ion number density ($n_i$) & 8.6 $\pm$ 0.4 cm$^{-3}$ \\
\Alfven\ speed ($v_A$) & 40 $\pm$ 2 km s$^{-1}$ \\
Perpendicular ion temperature ($T_{i\perp}$) & 7.5 $\pm$ 0.4 eV \\
Ion beta ($\beta_i$) & 1.1 $\pm$ 0.1 \\
Ion temperature anisotropy ($T_{i\perp}/T_{i\para}$) & 0.5 $\pm$ 0.2 \\
Ion gyroradius ($\rho_i$) & 74 $\pm$ 3 km \\
\Alfven\ ratio ($r_A$) & 0.72 $\pm$ 0.04 \\
\hline
\end{tabular}
\end{table}

The \Alfven\ ratio is the ratio of energy in the velocity $\mathbf{u}$ to the magnetic field in \Alfven\ units $\mathbf{b}$, and can be calculated spectrally, $r_A=E^u/E^b$, where $E^u$ and $E^b$ are the power spectra of $\mathbf{u}$ and $\mathbf{b}$. We calculate the average $r_A$ in the spacecraft frequency range $2\times 10^{-3}$ Hz to $1\times 10^{-2}$ Hz, which roughly corresponds to scales 36,000 km to 180,000 km under Taylor's hypothesis \citep{taylor38}. While this is at larger scales than the following anisotropy measurements, it is in the range where noise does not appear to dominate the velocity spectra. The value slightly less than unity that we obtain ($\approx$ 0.7) is consistent with previous measurements \citep[e.g.][]{matthaeus82a,marsch90a,podesta07a,bruno07,salem09}.

\begin{figure}
\begin{center}
\includegraphics[scale=0.5]{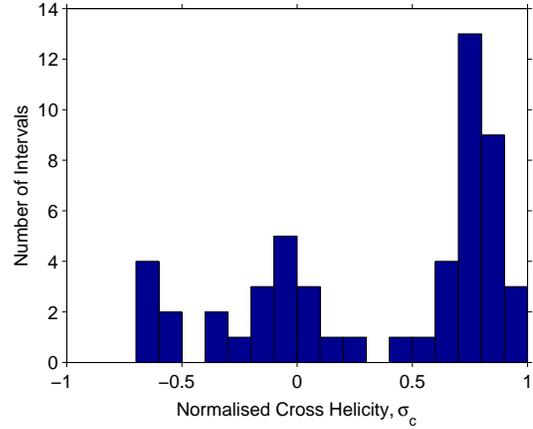}
\caption{\label{fig:nch}Histogram of normalised cross helicity, $\sigma_c$, for 53 of the the solar wind intervals in the spacecraft frequency range $2\times 10^{-3}$ Hz to $1\times 10^{-2}$ Hz.}
\end{center}
\end{figure}

We also calculate the normalised cross helicity, 
\begin{equation}
\label{eq:sigmac}
\sigma_c=\frac{E^+-E^-}{E^++E^-},
\end{equation}
where $E^+$ and $E^-$ are the power spectra of the Elsasser variables $\mathbf{z}^\pm=\mathbf{u}\pm\mathbf{b}$. The average value for each interval is calculated over the same range as the \Alfven\ ratio. The usual convention is used: the Elsasser variables are defined such that positive values of $\sigma_c$ correspond to \Alfvenic\ propagation away from the Sun. A histogram of $\sigma_c$ (Fig.~\ref{fig:nch}) shows a range of values with a non-Gaussian distribution: there is a large outward population ($\sigma_c > 0.5$), a balanced population ($\sigma_c \approx 0$), and a few inward intervals ($\sigma_c < -0.5$).

\subsection{Analysis Technique}
\label{sec:swanalysistechnique}

For each interval, pairs of points from the time series of the four spacecraft are used to calculate second-order structure functions at different angles to the local magnetic field, as described by \citet{chen10b}. The second-order structure function is defined as
\begin{equation}
\delta B^2_i (\mathbf{l}) = \left\langle |B_i (\mathbf{r}+\mathbf{l})-B_i(\mathbf{r})|^2\right\rangle, 
\end{equation}
where $B_i$ is the $i$th component of the magnetic field, $\mathbf{l}$ is the separation vector, and the angular brackets denote an ensemble average over positions $\mathbf{r}$. The local mean magnetic field at scale $\mathbf{l}$ is defined as 
\begin{equation}
\mathbf{B}_{\text{local}} = \frac{\mathbf{B}(\mathbf{r}+\mathbf{l})+\mathbf{B}(\mathbf{r})}{2}.
\end{equation}
We calculate the structure functions of the local perpendicular magnetic field component $\mathbf{B}_{\perp}$, which corresponds to the \Alfvenic\ fluctuations, at a variety of separations $\mathbf{l}$.

\begin{figure}
\begin{center}
\includegraphics[scale=0.5]{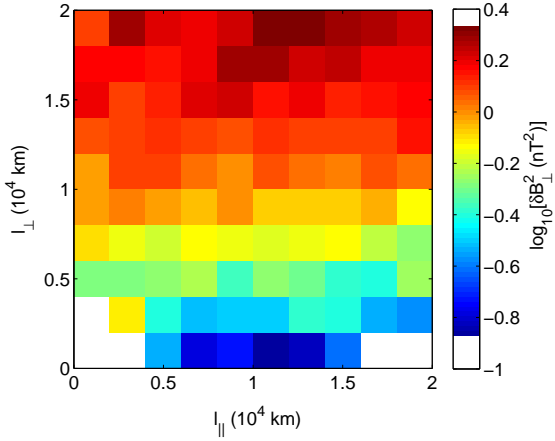}
\caption{\label{fig:s2ppsw}second-order structure function of the perpendicular magnetic field component for one of the 65 solar wind intervals as a function of parallel ($l_{\para}$) and perpendicular ($l_{\perp}$) separation.}
\end{center}
\end{figure}

The structure function values are binned according to scale parallel, $l_{\para}$, and perpendicular, $l_{\perp}$, to $\mathbf{B}_{\text{local}}$. Nine linearly spaced bins are used in each direction covering the range 2,000 km to 20,000 km, which is within, although towards the small scale end, of the inertial range. The result of this binning for one of the 65 intervals is shown in Fig.~\ref{fig:s2ppsw}. It is representative of the average behaviour, although in general each interval is more noisy and has less coverage than this. Most of the bin values in this figure are the average of a few thousand structure function values, although some (13\%) are of a few hundred. It can be seen that the contours are elongated in the field parallel direction, indicating that the eddies are anisotropic with $k_\perp>k_\para$.

The data is also binned according to scale $l$ and the angle $\theta_B$ between $\mathbf{l}$ and $\mathbf{B}_{\text{local}}$. Nine linearly spaced scale bins are used over the range 2,000 km to 20,000 km and nine linearly spaced angular bins are used between 0\degrees\ and 90\degrees. Straight lines, in log-log space, are then fitted to the structure functions over the full scale range for each $\theta_B$ bin and the power anisotropy is obtained by evaluating these fits at a scale of 10,000 km. The spectral index in each $\theta_B$ bin is found using the relation $\alpha =g+1$, where $-\alpha$ is the spectral index and $g$ is the structure function scaling exponent \citep{monin75}. This is similar to the work of \citet{osman09a}, except we bin the data with respect to the \emph{local} field direction, since this appears to be the relevant mean field for the fluctuations, and we use many more intervals.

\subsection{Magnetic Field Anisotropy}
\label{sec:mfa}

\begin{figure}
\begin{center}
\includegraphics[scale=0.53]{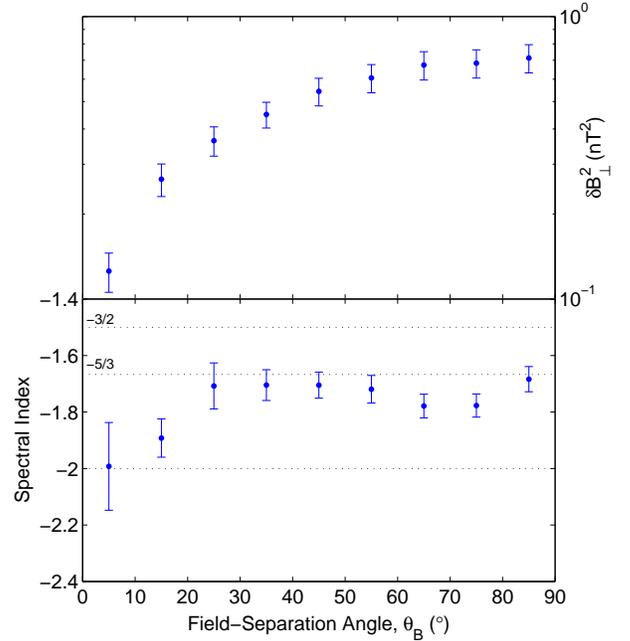}
\caption{\label{fig:swanisotropy}Power anisotropy (upper) and spectral index anisotropy (lower) of the perpendicular magnetic field component in inertial range turbulence in the slow solar wind. The power anisotropy is calculated at $l=10,000$ km. Spectral index values of $-3/2$, $-5/3$ and $-2$ are marked as dotted lines for reference.}
\end{center}
\end{figure}

The results, averaged over all 65 slow wind intervals, are shown in Fig.~\ref{fig:swanisotropy}, where the error bars are the standard error of the mean from averaging the intervals. They are similar to previous single spacecraft observations in the fast wind that show that power increases with $\theta_B$ and that the spectral index varies from $-2$ at small angles to between $-5/3$ and $-3/2$ at large angles \citep{horbury08,podesta09a,luo10,wicks10a,wicks11}. This scale-dependent anisotropy, therefore, has now been seen in both fast and slow wind using two different measurement techniques. The power anisotropy is consistent with eddies elongated along the local magnetic field direction and wavevector anisotropy of the form $k_{\perp}>k_{\para}$ \citep{chen10a}. 

The $-2$ scaling at small $\theta_B$ is consistent with both the theories of \citet{goldreich95} and \citet{boldyrev06}, which describe critically balanced \Alfvenic\ turbulence. It has been suggested \citep[e.g.][]{galtier10}, however, that the parallel scaling of $-2$ may be due to discontinuities in the data. Using the same technique at smaller scales in the dissipation range, the $-2$ scaling is not seen \citep{chen10b} because the physics of the turbulence is different at these scales. This suggests that the $-2$ scaling in the inertial range seen here is in fact due to the properties of the turbulence and not unrelated discontinuities. 

The perpendicular spectral index that we obtain here (for $20\degrees<\theta_B< 90\degrees$) is closer to $-5/3$ than $-3/2$. This agrees with the prediction of \citet{goldreich95}, rather than \citet{boldyrev06}. Both of these theories, however, apply to balanced turbulence, i.e., $\sigma_c=0$. As can be seen from Fig.~\ref{fig:nch}, many of the intervals have large $\sigma_c$. This is common in the solar wind and various theories of imbalanced MHD turbulence have been proposed \cite[e.g.][]{lithwick07,chandran08,beresnyak08,perez09,podesta10c}. Differentiating between these, however, is beyond the scope of this paper [see \citet{wicks11} for a recent observational test of these theories].

In the next section, we apply a similar analysis to RMHD simulations. This enables a direct comparison to be made between turbulence in the solar wind and in numerical simulations.

\section{Reduced MHD Simulations}
\label{sec:sim}

\subsection{Simulation Description}
\label{sec:simdesc}

The RMHD equations, originally derived by \citet{strauss76}, have been used previously to simulate various aspects of MHD turbulence \citep[e.g.][]{perez08,perez09,beresnyak11}. They can be written in Elsasser potentials \citep{schekochihin09}:
\begin{multline}
\label{eq:rmhd}
\frac{\partial}{\partial t} \nabla^2_\perp \zeta^\pm \mp v_A \frac{\partial}{\partial z} \nabla^2_\perp \zeta^\pm \\
= - \frac{1}{2} \left(\{\zeta^+ , \nabla^2_\perp \zeta^- \} + \{\zeta^- , \nabla^2_\perp \zeta^+ \} \mp \nabla^2_\perp \{\zeta^+ , \zeta^-\}\right),
\end{multline}
where $\left\{A,B\right\} = \mathbf{\hat{z}}\cdot\left(\nabla_\perp A\times\nabla_\perp B\right)$, $\mathbf{\hat{z}}$ is the global mean field direction, $v_A$ is the \Alfven\ speed and the Elsasser potentials are defined via $\delta\mathbf{z}_\perp^\pm=\delta\mathbf{u}_\perp\pm\delta\mathbf{b}_\perp=\mathbf{\hat{z}}\times\nabla_\perp\zeta^\pm$.

Equations (\ref{eq:rmhd}) contain only the perpendicular fluctuations and are, therefore, suitable for simulating \Alfvenic\ turbulence. They are also more efficient to simulate than MHD, since they involve only two scalar fields. Although originally derived from MHD, it has been shown that RMHD holds for a collisionless plasma such as the solar wind and may, therefore, be more generally applicable \citep{schekochihin09}. The RMHD derivation assumes anisotropy ($k_{\perp} \gg k_{\para}$) and a strong mean field ($B_0 \gg \delta \mathbf{B}_{\perp}$), both of which are observed at the smallest scales of the solar wind inertial range. 

The simulation reported here solves the RMHD equations in a triply periodic cube of size $(2\pi)^3$ with a resolution of $512^3$. The \Alfven\ speed is set to $v_A=1$ (making the \Alfven\ crossing time $2\pi$). It can be seen from equations (\ref{eq:rmhd}) that if the \Alfven\ speed is scaled by a factor $R$ and the $z$ coordinate, which is the mean field direction, is also scaled by $R$, the equations remain identical. This means that a given simulation corresponds to all values of $R$, and therefore all values of $\delta\mathbf{B}_{\perp}/B_0$ if the box is also stretched in the $z$ direction. The units of length in the perpendicular and parallel directions are independent of each other because the anisotropy is formally infinite and the fluctuation level is infinitely small under the RMHD asymptotic expansion. Different values of $R$ can be chosen, setting the anisotropy and fluctuation level so that the same simulation can be compared to a variety of real world situations.

The equations are solved pseudospectrally in $x$ and $y$, and using a centred finite difference scheme in $z$. The time step is chosen so that the Courant numbers based on both the fluctuation amplitude and the \Alfven\ speed are much less than unity. With dissipation and forcing terms the equations are
\begin{multline}
\label{eq:rmhdvisc}
\frac{\partial}{\partial t} \nabla^2_\perp \zeta^\pm \mp v_A \frac{\partial}{\partial z} \nabla^2_\perp \zeta^\pm \\
= - \frac{1}{2} \left(\{\zeta^+ , \nabla^2_\perp \zeta^- \} + \{\zeta^- , \nabla^2_\perp \zeta^+ \} \mp \nabla^2_\perp \{\zeta^+ , \zeta^-\}\right) \\
+\nu \nabla_\perp^8 (\nabla_\perp^2 \zeta^\pm)+\nu_z \nabla_z^2 (\nabla_\perp^2 \zeta^\pm)+f^\pm,
\end{multline}
where $\nu=5\times10^{-15}$ and $\nu_z=1\times10^{-4}$ are the viscosity coefficients and $f^\pm$ is the forcing term. In the $x$ and $y$ directions, a 4th order hyperviscosity dissipation term is used, while in the $z$ direction a very small Laplacian viscosity is added to prevent the high $k_z$ modes becoming unstable. Hyperviscosity is used so that the inertial range covers a wide enough range of scales to measure accurate scalings. The magnetic Prandtl number is $\text{Pr}_\text{m}=1$ and the initial conditions are a straight mean field with no fluctuations: $\mathbf{b}(\mathbf{r},t=0)=\mathbf{\hat{z}}$ and $\mathbf{u}(\mathbf{r},t=0)=0$.

The simulation is initially forced on large scales ($k_\perp=1,2$ and $k_z=1$) with Gaussian white noise forcing $f^\pm$, i.e. the random forcing amplitude is refreshed at each time step. This means that the input power can be controlled; it is set to unity in the code units to produce strong turbulence. We choose to force only the velocity to match possible sources of solar wind forcing, such as velocity shears or large scale \Alfven\ waves, so $f^+=f^-$ at all times. We do not force the magnetic field since there is no known mechanism of breaking magnetic flux conservation at large scales. After a while, the forcing is removed and the simulation is left to freely decay.

\begin{figure}
\begin{center}
\includegraphics[scale=0.53]{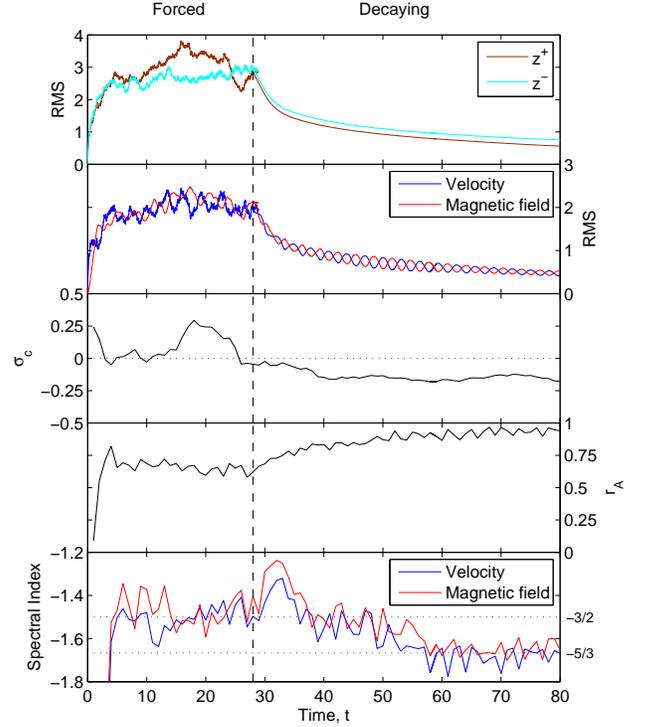}
\caption{\label{fig:timeseries}Time series of simulation parameters: RMS variables, normalised cross helicity $\sigma_c$, \Alfven\ ratio $r_A$ and Fourier perpendicular spectral indices. For $t\leq28$ the simulation is forced and for $t>28$ it is decaying. Spectral index values of $-3/2$ and $-5/3$ are marked as dotted lines for reference in the lower panel.}
\end{center}
\end{figure}

A time series of various simulation parameters is shown in Fig.~\ref{fig:timeseries}. After the simulation begins, the values take a few time units to settle down, which is roughly the turnover time of the largest eddies. The transition between the forced and decaying periods of the simulation can be seen by the change in behaviour of all the quantities at $t=28$, marked by the dashed line. The top two panels show the root mean square (RMS) values of the Elsasser variables, velocity and magnetic field. Their values up to $t=28$ are determined by the forcing power and after $t=28$ by the decay of the turbulence. One noticeable feature is the oscillation in the velocity and magnetic field RMS values with a period $\approx 2\pi$. This is most likely due to large scale \Alfven\ waves, also seen by \citet{bigot08a}, which should not significantly affect the average inertial range measurements.

The normalised cross helicity $\sigma_c$ in the third panel is calculated spectrally [equation (\ref{eq:sigmac})], as was done for the solar wind intervals in Section \ref{sec:swintervals}, and averaged over the range $7\leq k_\perp\leq 33$. During the forced period, $\sigma_c$ fluctuates above and below 0. When the forcing is removed, $|\sigma_c|$ increases, as expected from dynamic alignment theory \citep{dobrowolny80}. The increase is fairly slow: $|\sigma_c|$ changes from $0.045$ at $t=28$ to $0.13$ at $t=78$, which is consistent with previous decaying simulations \citep[e.g.][]{grappin82,matthaeus83,pouquet86,oughton94}.

The \Alfven\ ratio $r_A$ is shown in the fourth panel, calculated over the same range as $\sigma_c$. During the forced period $4\leq t\leq 28$, its mean value is $r_A \approx 0.66$, which is close to the solar wind observations (Table \ref{tab:parameters}). As the turbulence decays, $r_A$ grows and approaches unity; this equipartition of energy is expected for MHD turbulence \citep{kraichnan65}. We note that the opposite effect is seen in simulations without a strong mean field \citep[e.g.][]{oughton94,biskamp99b}, in which the \Alfven\ ratio decreases away from unity as the energy decays. The fact that the equipartition occurs only in the decaying period of our RMHD simulation, while solar wind observations show $r_A<1$ \citep[e.g.][]{matthaeus82a,marsch90a,podesta07a,bruno07,salem09}, suggests that solar wind turbulence may be better described by a forced model.

The perpendicular spectral indices for the velocity and magnetic field are shown in the lower panel of Fig.~\ref{fig:timeseries}. They are calculated from the gradients of the best fit lines to the perpendicular energy spectrum in log-log space over the range $7\leq k_\perp\leq 33$ every time unit. The perpendicular spectrum is calculated as the sum of the energy in modes nearest to $k_\perp=\sqrt{k_x^2+k_y^2}$ for integer values of $k_\perp$. During the forced period, the spectral indices are closer to $-3/2$ than $-5/3$, in agreement with previous results \citep{maron01,muller03,muller05,mason08,perez08,grappin10}. When the forcing is removed, however, they gradually steepen and appear to reach a steady value of $-5/3$ from $t=58$ onwards.

\begin{figure}
\begin{center}
\includegraphics[scale=0.53]{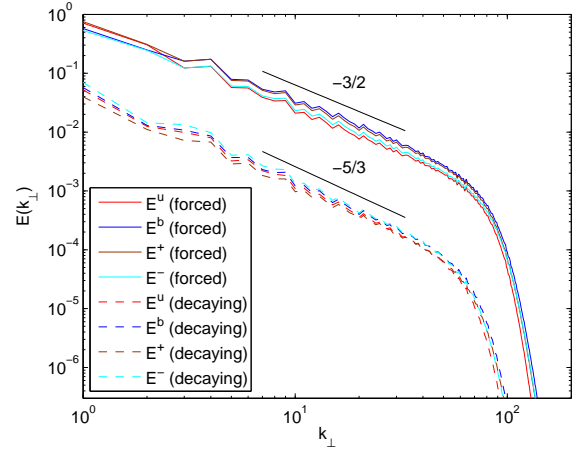}
\caption{\label{fig:spectra}Perpendicular energy spectra of velocity ($E^u$), magnetic field ($E^b$), and Elsasser variables ($E^\pm$) in the forced and decaying periods of the simulation. Slopes of $-5/3$ and $-3/2$ are given for reference.}
\end{center}
\end{figure}

In the following analysis, we investigate the anisotropic scaling in the forced period $4\leq t\leq 28$, and the decaying period $58\leq t\leq 78$. We assume that in each of these periods the turbulence is stationary and we can perform time averages over them. The averaged energy spectra are shown in Fig.~\ref{fig:spectra}. Before averaging, the decaying spectra are normalised so that the average energy over the range $7\leq k_\perp\leq 33$ for each is the same as that at $t=58$. Gradients of $-5/3$ and $-3/2$ are given for reference, although it is hard to tell the difference between these visually. It can be seen that for $7\leq k_\perp\leq 33$ there are well defined power laws in all of the spectra. It has been suggested \cite[e.g.][]{perez10b} that the use of hyperviscosity may increase the bottleneck effect, altering the scaling. The spectra in Fig.~\ref{fig:spectra} do not, however, display the increase of energy at small scales that is associated with the bottleneck effect and is seen in some MHD simulations \citep[e.g.][]{cho00,beresnyak11}. In the next section, we measure the anisotropic scaling using structure functions, which are expected to be less susceptible to the bottleneck effect than Fourier spectra \citep{dobler03}.

\subsection{Analysis Technique}

The technique we use to analyse the simulation data is similar to that used in Section \ref{sec:swanalysistechnique}, with modifications to account for the simulation geometry. Firstly, the scaling factor $R$, which should be larger than unity for the RMHD equations to be valid, is chosen. Here, we set $R=4$, which is a compromise between typical solar wind wavevector anisotropies $k_\perp/k_\para$ of between 2 and 3 and typical $\delta\mathbf{B}_\perp/B_0$ values of 0.1 (calculated from the data in Section \ref{sec:mfa}). This means that the simulation, which was solved in a $(2\pi)^3$ box, is now stretched to have a size $(2\pi)^2\times8\pi$ and the \Alfven\ speed is set to 4.

\begin{figure}
\begin{center}
\includegraphics[scale=0.5]{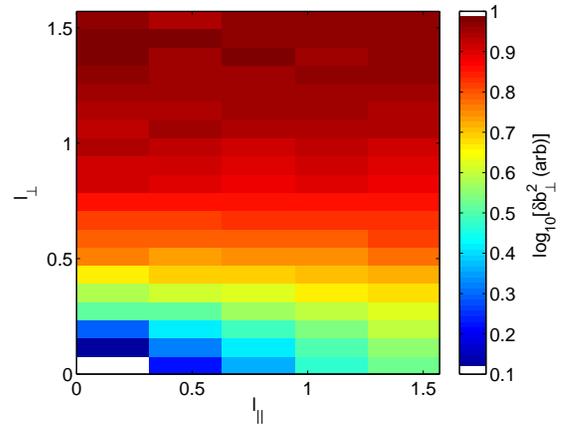}
\caption{\label{fig:s2ppsim}second-order structure function of the perpendicular magnetic field component for one of the snapshots ($t=28$) in the forced simulation as a function of parallel ($l_{\para}$) and perpendicular ($l_{\perp}$) separation.}
\end{center}
\end{figure}

For a particular snapshot in time, many pairs of points in the simulation box are picked at random. The second-order structure function values of the local perpendicular velocity and magnetic field components are calculated and binned, as in Section \ref{sec:swanalysistechnique}. The structure function of the magnetic field binned with respect to $l_{\para}$ and $l_{\perp}$ at $t=28$ is shown in Fig.~\ref{fig:s2ppsim}. There are on average $10^4$ structure function values in each bin. The structure function in Fig.~\ref{fig:s2ppsim} is representative of the general shape of the velocity and magnetic field structure functions in both the forced and decaying periods of the simulation. Similarly to the solar wind (Fig.~\ref{fig:s2ppsw}) and previous simulations \citep{cho00}, the contours are elongated in the parallel direction.

In the range $0.35 \leq l \leq 1.3$, which corresponds approximately to $5 \leq k \leq 18$, the structure functions are approximately power laws and we assume this to be the inertial range of the simulation. The spectral indices and the power anisotropy (calculated at $l=0.8$) are found from the best fit lines to the data binned with respect to $l$ and $\theta_B$ in this range. This is done for snapshots separated by 2 time units, giving 13 snapshots for the forced period and 11 for the decaying period.

\begin{figure*}
\begin{center}
\subfigure{\includegraphics[scale=0.53]{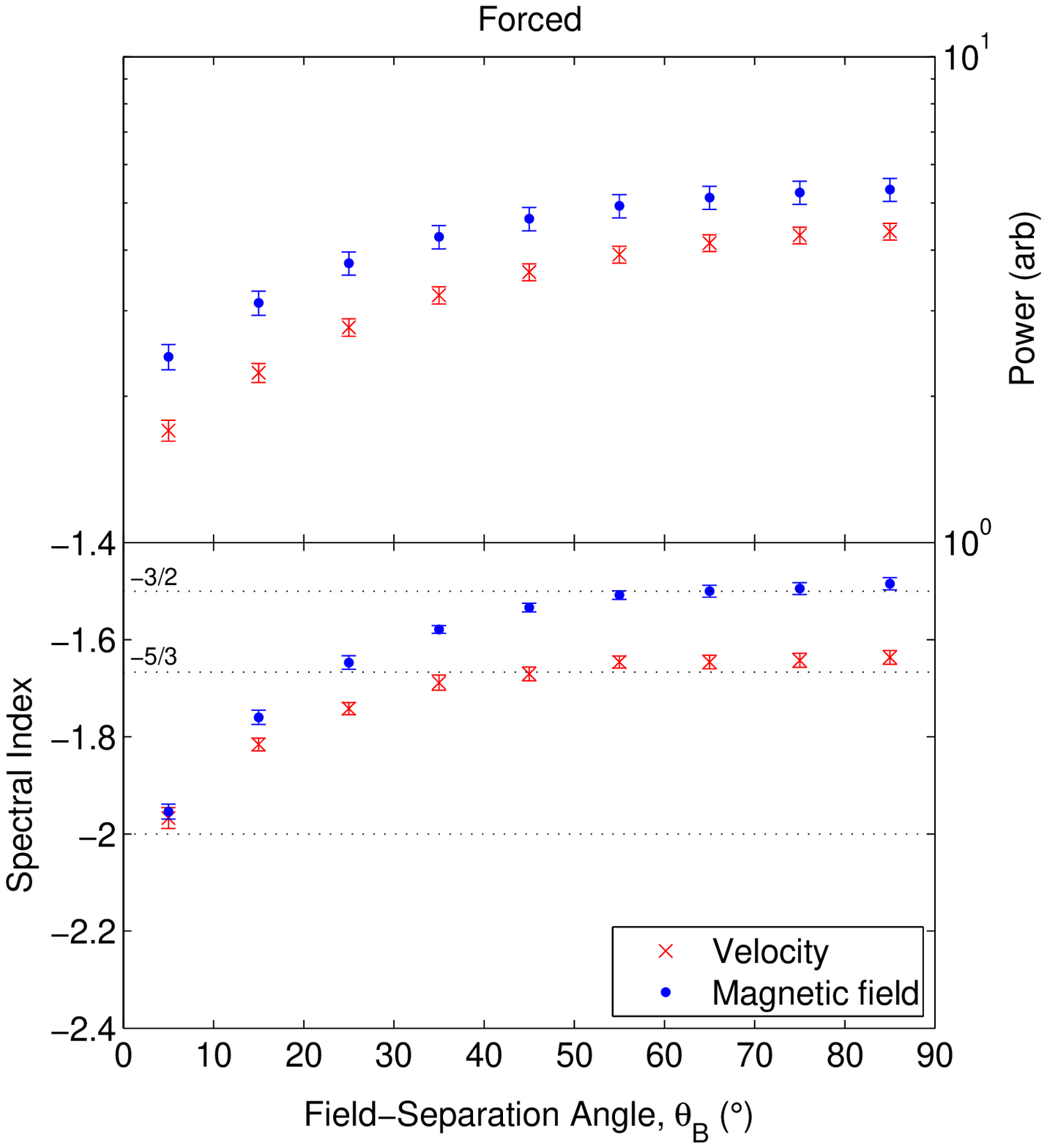}}
\subfigure{\includegraphics[scale=0.53]{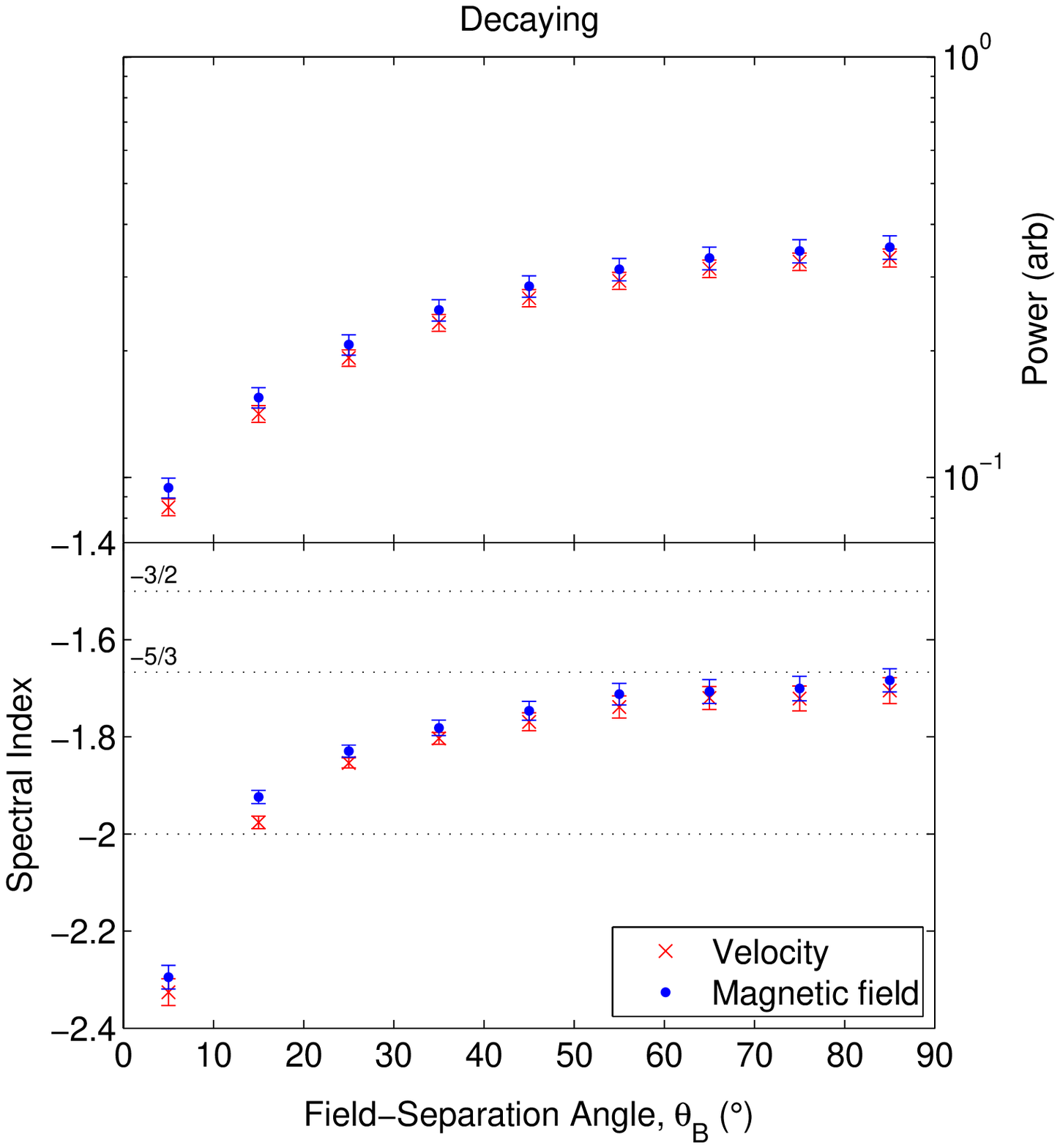}}
\caption{\label{fig:simanisotropy}Power anisotropy (upper panels) and spectral index anisotropy (lower panels) of the velocity and magnetic field in the forced (left) and decaying (right) simulations. The power anisotropy is calculated at $l=0.8$. Spectral index values of $-3/2$, $-5/3$ and $-2$ are marked as dotted lines for reference.}
\end{center}
\end{figure*}

\subsection{Power and Spectral Index Anisotropy}

The power and spectral index anisotropy for the velocity and magnetic field are shown in Fig.~\ref{fig:simanisotropy}. The error bars are the standard error of the mean from averaging the results of the snapshots in each period. In both cases, the power increases with angle to the local magnetic field $\theta_B$, as it does in the solar wind. For the forced case, the overall power in the magnetic field is larger than that in the velocity, whereas in the decaying case they are similar. This is consistent with our previous discussion of the \Alfven\ ratio being $r_A<1$ in the forced case and $r_A\approx 1$ in the decaying case. It is also interesting to note that these curves are qualitatively similar in shape. No prediction for this shape has yet been made based on critical balance theory.

In the forced case, there is a difference between the spectral indices of the velocity and magnetic field. The velocity spectral index varies from $-2$ at small $\theta_B$ to $-5/3$ at $\theta_B$ close to $90\degrees$. The magnetic field spectral index is also $-2$ at small $\theta_B$ but is less steep at larger $\theta_B$, having a value close to $-3/2$. The fact that both are steep at small angles shows that the turbulence is anisotropic and the $-2$ scaling is consistent with the critical balance theories of both \citet{goldreich95} and \citet{boldyrev06}. The difference at large angles, however, is unexpected, since theories of \Alfvenic\ turbulence predict that both fields scale in the same way.

For the decaying case, both fields show similar scaling. The spectral index is close to $-5/3$ for $\theta_B$ close to $90\degrees$ and much steeper at small $\theta_B$: $-2.33 \pm 0.03$ for velocity and $-2.30 \pm 0.03$ for the magnetic field. Again, the steepening at low $\theta_B$ shows the turbulence is anisotropic, although the spectra are steeper than the critical balance prediction of $-2$. One possible explanation for the steep parallel scaling is that the turbulence may be transitioning to the weak regime, in which there is not thought to be a parallel cascade \citep{goldreich97,galtier00}. \citet{perez08} observed the perpendicular spectral index steepening as the turbulence became weaker, and we may be observing a similar effect for the parallel index. In a different run (not shown here) that was forced less strongly, we observed overall steeper spectral indices at all angles. The perpendicular spectral index, however, seems to remain at $-5/3$ for many turnover times in the run here (Fig.~\ref{fig:timeseries}), rather than dropping to $-2$ as expected for weak turbulence.

We now compare the spectral indices obtained through the structure function technique to the Fourier indices. The time series of the global perpendicular Fourier indices are shown in the lower panel of Fig.~\ref{fig:timeseries}. It can be seen that both fields have spectral indices close to $-3/2$ during the forced period and then after a transition, reach a value of $-5/3$ in the decaying period. The mean values are $-1.51 \pm 0.01$ for the velocity and $-1.47 \pm 0.01$ for the magnetic field in the forced period and $-1.69 \pm 0.01$ for the velocity and $-1.653 \pm 0.007$ for the magnetic field in the decaying period. These are consistent with the perpendicular spectral indices measured using structure functions, except for velocity in the forced period, which is close to $-5/3$. It is possible that this difference is caused by the forcing, which is localised at large scales in Fourier space, but may affect the structure function, which mixes small and large scale information \citep{davidson05}.

The results we obtain here are broadly consistent with previous simulations. Wavevector anisotropy of the form $k_{\perp}>k_{\para}$ has been observed previously \citep[e.g.][]{shebalin83,oughton94,matthaeus96,milano01}. In particular, \citet{cho00} observed a difference in anisotropic scaling between the velocity and magnetic field in their forced simulations. When the mean field was of a similar strength to the RMS fluctuations they obtained $k_{\para}\sim k_{\perp}^{0.7}$ for velocity but $k_{\para}\sim k_{\perp}^{0.5}$ for the magnetic field.

We now compare the simulation and solar wind results. Firstly we note that both sets of results are qualitatively similar. Power at a fixed scale is anisotropic and increases as $\theta_B$ increases. All spectral index curves are anisotropic and steepen at small $\theta_B$ as predicted by critical balance theories. The main difference between the solar wind and simulations is the value of the perpendicular spectral index. For the magnetic field, we observe $-5/3$ in the solar wind and the decaying simulation but $-3/2$ in the forced simulation. Values close to both $-5/3$ and $-3/2$ have been observed previously in the solar wind \citep{horbury08,podesta09a,luo10,wicks10a,wicks11}. In both our forced and decaying simulations, the velocity has a perpendicular spectral index of $-5/3$. Solar wind measurements, however, suggest that it is closer to $-3/2$ \citep{mangeney01,podesta07a,tessein09,salem09,wicks11}. These differences in perpendicular spectral index remain an unsolved problem.

\section{Local vs Global Mean Field}
\label{sec:localvsglobal}

\begin{figure}
\begin{center}
\includegraphics[scale=0.53]{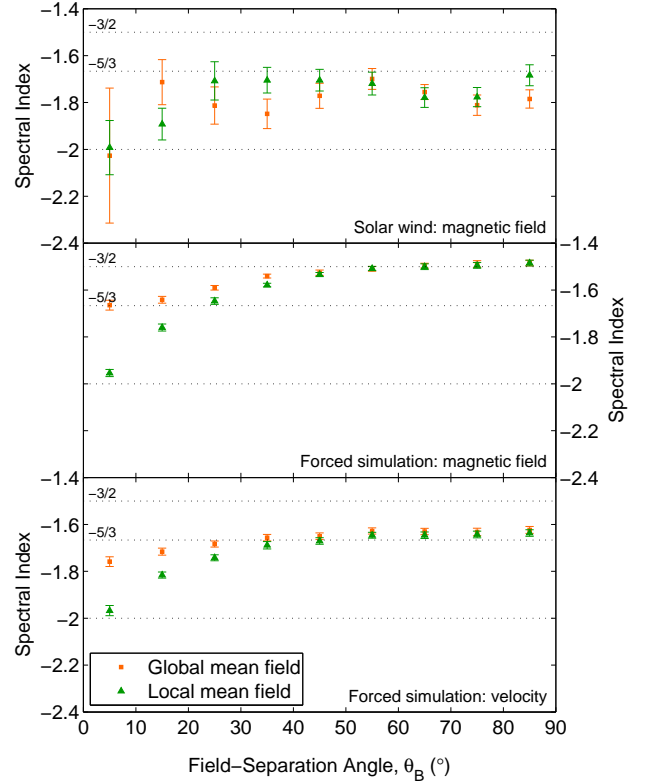}
\caption{\label{fig:localvsglobal}Spectral index anisotropy of magnetic field in the solar wind (upper), magnetic field in the forced simulation (centre) and velocity in the forced simulation (lower) using the local and global mean field methods. Spectral index values of $-3/2$, $-5/3$ and $-2$ are marked as dotted lines for reference.}
\end{center}
\end{figure}

In this section, we investigate the difference between using the local and global mean magnetic field to define the parallel and perpendicular directions. Fig.~\ref{fig:localvsglobal} shows the spectral index anisotropy for the solar wind magnetic field in the upper panel, the forced simulation magnetic field in the centre panel and the forced simulation velocity in the lower panel. In each case, the results obtained using the local mean field are shown in green and those obtained using the global mean in orange. In the solar wind, the global mean field results are obtained by binning the structure function values according to their separation direction with respect to the average field of each interval. In the simulation, they are obtained by binning with respect to the average field over the whole simulation box (the $z$ direction).

In the solar wind, the results are similar for both local and global mean field methods, except for the error bar on the parallel scaling ($0\degrees<\theta_B<10\degrees$). The error is larger in the global mean field case, which is partly due to the fact that there are fewer intervals where the global mean field is parallel to any separation vectors. This scaling, therefore, is less reliable and the error bar indicates that the data is marginally consistent with an isotropic spectral index with respect to the global mean field. To within errors, these results are not inconsistent with those of \citet{tessein09}, which show that the spectral index is isotropic when measured with respect to the global mean field.

In the forced simulation, it can be seen that when measured with respect to the global mean field, the spectral indices of both the magnetic field and velocity are much less anisotropic than when they are measured with respect to the local mean field, e.g. in the velocity at small $\theta_B$ the spectral index is $-1.76 \pm 0.02$ using the global mean field compared to $-1.97 \pm 0.02$ using the local mean field. This is because the magnetic field fluctuations are large enough that the local mean field direction seen by an eddy is not the same as the global mean field direction. If the fluctuations are in critical balance, the angle between the local and global mean fields is $\delta\mathbf{B}_\perp/B_0\approx k_\para/k_\perp$. This suggests that when using the global mean field, the parallel scaling cannot be correctly distinguished from the perpendicular scaling, even for small $\delta\mathbf{B}_\perp/B_0$, because the angle of measurement to the local mean field needs to be less than $k_\para/k_\perp$.

This interpretation is in agreement with previous solar wind studies that have used local and global mean field methods. Those that use the global mean field method do not detect spectral index anisotropy \citep{sari76,tessein09} and those that use a local mean field method do detect it \citep{horbury08,podesta09a,luo10,wicks10a,wicks11}. A similar situation is also seen in simulations, where scaling anisotropy is detected when a local mean field is used \citep{cho00,maron01} but not when a global mean field is used \citep{grappin10}. Here, we have shown that when keeping all other parameters constant, it is indeed the use of the global or local mean field that determines whether the anisotropic scaling is measured. It seems, therefore, that the \Alfvenic\ fluctuations, both in solar wind turbulence and forced RMHD turbulence simulations, are more sensitive to the local mean field at the scale of the fluctuations than the global large scale field.

In the decaying simulation (not shown in Fig.~\ref{fig:localvsglobal}), the local and global mean field methods are much more similar, with the parallel scaling being steeper than $-2$ in all cases. One possible reason for this is that the scale separation between the global mean field and the fluctuations is not large, meaning that the global and local mean fields are similar. This, combined with the smaller fluctuation amplitudes in the decaying simulation, could account for the observed behaviour. This could be tested by performing a decaying simulation with a larger inertial range.

\section{Summary and Conclusions}
\label{sec:conc}

In this paper, we measure the power and spectral index anisotropy of \Alfvenic\ turbulence in the solar wind and RMHD simulations using second-order structure functions. The analysis technique is essentially the same for both, allowing us to make a direct comparison. In the slow solar wind, we find that the magnetic field power and spectral index are anisotropic with respect to the local magnetic field direction. This anisotropy has now been seen by several different methods in both fast and slow wind. In both forced and decaying simulations we also find that the power and spectral index are anisotropic in both the velocity and magnetic field.

In the solar wind, the perpendicular spectral index of the magnetic field is close to $-5/3$, in agreement with the theory of \citet{goldreich95}. In the forced simulation, the perpendicular spectral indices are close to $-5/3$ for velocity and $-3/2$ for the magnetic field. We are not aware of any theory that can account for this difference, although it may be caused by the velocity forcing. In the decaying simulation, the perpendicular spectral index is close to $-5/3$ for both the velocity and magnetic field. In all cases, the spectral index steepens at small angles to the magnetic field. The parallel scaling obtained in the solar wind and forced simulations is close to $-2$, which agrees with the theories based on critical balance of both \citet{goldreich95} and \citet{boldyrev06}. The parallel spectral indices in the decaying simulation are $-2.33 \pm 0.03$ for the velocity and $-2.30 \pm 0.03$ for the magnetic field, which are steeper than the critical balance predictions.

We also find that when measuring the anisotropy of the fluctuations in the forced simulation with respect to the global magnetic field, rather than the local mean field, the spectral indices are much less anisotropic. This is expected for critically balanced turbulence and is also consistent with previous solar wind and simulation results: those that used the local mean field saw anisotropic scaling and those that used the global mean field did not.

\section*{Acknowledgments}

This work was supported by STFC and the Leverhulme Trust Network for Magnetized Plasma Turbulence. FGM and CIS data were obtained from the Cluster Active Archive. The simulations were carried out using resources at the Texas Advanced Computing Center. We acknowledge useful discussions with R.~Wicks and helpful comments from an anonymous referee.

\end{document}